\documentclass[12pt,preprint]{aastex} 

\renewcommand{\deg}{\ifmmode^\circ\else$^\circ$\fi}

\newcommand{\arcm}{\hbox{$^\prime$}}
\newcommand{\arcs}{\hbox{$^{\prime\prime}$}}

\newcommand{\um}{\hbox{$\mu$m}}
\newcommand{\xx}{$\times$}

\shorttitle{Mid-IR Properties of X-ray Sources} 
\shortauthors{Gorjian et al.}

\begin{document}

\title{The Mid-Infrared Properties of X-ray Sources }

\author{Gorjian, V., Brodwin, M.}
\affil{MS 169-327, Jet Propulsion Laboratory, California Institute of Technology, 4800 Oak Grove Dr., Pasadena, CA 91109}

\author{Kochanek, C.S.}
\affil{Department of Astronomy, Ohio State University, Columbus, OH 43210}

\author{Murray, S.}
\affil{Harvard-Smithsonian Center for Astrophysics, 60 Garden Street, Cambridge, MA 02138}

\author{Stern, D.} 
\affil{MS 169-327, Jet Propulsion Laboratory, California Institute of Technology, 4800 Oak Grove Dr., Pasadena, CA 91109}

\author{Brand, K.}
\affil{National Optical Astronomy Observatory, 950 North Cherry Avenue, Tucson, AZ 85726}

\author{Eisenhardt, P. R.} 
\affil{MS 169-327, Jet Propulsion Laboratory, California Institute of Technology, 4800 Oak Grove Dr., Pasadena, CA 91109}

\author{Ashby, M.L.N., Barmby, P.}
\affil{Harvard-Smithsonian Center for Astrophysics, 60 Garden Street, Cambridge, MA 02138}

\author{Brown, M. J. I.}
\affil{School of Physics, Monash University, Clayton Vic 3800, Australia}

\author{Dey, A.}
\affil{National Optical Astronomy Observatory, 950 North Cherry Avenue, Tucson, AZ 85726}

\author{Forman, W.}
\affil{Harvard-Smithsonian Center for Astrophysics, 60 Garden Street, Cambridge, MA 02138}

\author{Jannuzi, B.T.}
\affil{National Optical Astronomy Observatory, 950 North Cherry Avenue, Tucson, AZ 85726}

\author{Jones, C., Kenter, A.T., Pahre, M.A.}
\affil{Harvard-Smithsonian Center for Astrophysics, 60 Garden Street, Cambridge, MA 02138}

\author{Shields, J.C.}
\affil{Department of Physics and Astronomy, Ohio University, Athens, OH 45701}

\author{Werner, M.W.} 
\affil{MS 264-767, Jet Propulsion Laboratory, California Institute of Technology, 4800 Oak Grove Dr., Pasadena, CA 91109}

\author{Willner, S.P.}
\affil{Harvard-Smithsonian Center for Astrophysics, 60 Garden Street, Cambridge, MA 02138}

\email{varoujan.gorjian@jpl.nasa.gov}

\begin{abstract}

We combine the results of the {\it Spitzer} IRAC Shallow Survey and the {\it Chandra} XBo\"otes Survey
of the 8.5 square degrees Bo\"otes field of the NOAO Deep Wide-Field Survey to produce the 
largest comparision of mid-IR and X-ray sources to date.
The comparison is limited to sources with X-ray fluxes $>$8\xx10$^{-15}$ erg cm$^{-2}$s$^{-1}$ 
in the 0.5-7.0 keV range and mid-IR sources with 3.6 \um\ fluxes brighter than
18.4 mag (12.3~$\mu$Jy).  In this most sensitive IRAC band, 85\% of the 3086
X-ray sources have mid-IR counterparts at an 80\% confidence level based on a Bayesian matching technique. Only 2.5\% of the sample have no IRAC counterpart at all based on visual inspection. Even for a smaller but a significantly deeper {\it Chandra} survey in the same field, the IRAC Shallow Survey recovers most of the X-ray sources. A majority (65\%) of the {\it Chandra} sources 
detected in all four IRAC bands occupy a well-defined region of IRAC [3.6] - [4.5] vs [5.8] - [8.0]
color-color space. These X-ray sources are likely infrared luminous, unobscured  {\it type I} AGN with little
mid-infrared flux contributed by the AGN host galaxy. Of the remaining {\it Chandra} sources, 
most are lower luminosity {\it type I} and {\it type II} AGN whose mid-IR emission is dominated by the host galaxy, while approximately  5\% are either Galactic stars or very local galaxies.

\end{abstract}

\section{Introduction}

The Cosmic Infrared Background (CIB) represents a large fraction of the energy generated during the history of the universe.  
The two main contributers to the CIB are fusion processes within stars and accretion processes on to black holes 
\citep[see the review by][]{hauser01}.  The contribution to the CIB from accretion is less well constrained than that 
from stars because AGN surveys have always suffered from systematics and incompleteness. Optical and soft X-ray surveys 
tend to miss dust- or gas-obscured AGN, while surveys at wavelengths insensitive to absorption either include
only small sub-populations of AGN (radio) or are very limited by flux or survey area limitations (hard X-ray, infrared).  
With the advent of the {\it Chandra X-ray Observatory} \citep{weisskopf96} and the {\it Spitzer Space Telescope} \citep{werner04} the tools became available to do the wide-field imaging surveys needed to determine the relative contribution of stars and AGN to the CIB throughout the history of the universe.

In this paper we examine the mid-IR colors of a large sample of X-ray sources in the $\sim 9$~square degree Bo\"otes field of 
the NOAO Deep Wide Field Survey \citep[NDWFS;][]{jannuzi99}, a ground based survey from the optical to the near-IR (B$_W$RIJK$_S$). 
The X-ray sources were detected in the {\it Chandra} XBo\"otes survey of the field \citep{murray04, kenter05, brand06, hickox07}, and then matched to the 
mid-IR sources found in the {\it Spitzer} IRAC Shallow Survey of the field \citep{eisenhardt04}. 
\citet{eisenhardt04} pointed out that in a [3.6] - [4.5] vs [5.8] - [8.0] color-color diagram there was a distinct
branch of point-like 3.6 \um\ sources that were probably AGN.   This was confirmed by \citet{stern05} using spectroscopy
of nearly 10,000 of the Shallow Survey sources including 681 AGN from the AGN and Galaxy Evolution Survey (AGES, Kochanek et al. in 
preparation).  \citet{lacy04} found a similar color grouping in the {\it Spitzer} First Look Survey based on 54 optically identified quasars.  In this paper we characterize
the mid-IR properties of the full {\it Chandra} XBo\"otes sample.   There have been several previous surveys 
comparing X-ray and mid-IR sources \citep[e.g.][]{fadda02, alonso04,franceschini05,barmby06}, but our present survey is significantly
larger.  Our objectives are to characterize the mid-IR color distributions of X-ray sources and the efficiency with which 
we can detect X-ray sources in the mid-IR.  In \S2 we describe the X-ray and mid-IR observations and our method for 
matching the two catalogs. In \S3 we compare the two samples, and we summarize the results in \S4.

\section{Observations}

The XBo\"otes survey \citep{murray04} imaged roughly 8.5 square degrees of the Bo\"otes NDWFS using the Advanced CCD Imaging 
Spectrometer (ACIS) instrument on {\it Chandra} in a series of 126 short (5 ks) observations. The resulting limiting 
flux was $\sim$10$^{-3}$ cts s$^{-1}$ corresponding to $\sim$8\xx10$^{-15}$ erg cm$^{-2}$s$^{-1}$ in the 0.5-7.0 keV range. 
We use a catalog of 3442 sources with 4 or more counts that has a spurious detection rate of approximately 
1\% \citep{kenter05}.  Matching the X-ray sources to their IRAC counterparts is complicated by the fact that  the positions 
of the X-ray sources have significant uncertainties that depend on the fluxes of the sources and their locations relative
to the {\it Chandra} optical axis.  The {\it Chandra} PSF is only 0\farcs6 FWHM on axis, but degrades quadratically
with off-axis distance, reaching 6\farcs0 when 10~arcmin off-axis.  We adopt an astrometric uncertainty appropriate to each source based on its off axis position, divided by the square root of the source counts, with a fixed minimum uncertainty of 1\farcs5 (90\% confidence). 

The IRAC Shallow Survey \citep{eisenhardt04} covers 8.5 square degrees of the NDWFS field with 3 or more 30 sec exposures 
per position resulting in 17,076 separate 5\arcm\xx5\arcm\ images in the four IRAC bands. 
Sources were identified and characterized using SExtractor v2.3.2 \citep{bertin96}, resulting 
in detections (within 6\arcs\ diameter apertures) of $\sim$ 270,000, 200,000, 27,000, and 26,000 sources brighter 
than the 5$\sigma$ detection limits of 18.4, 17.7, 15.5, and 14.5 Vega magnitudes (12.3, 14.9, 72.3, 102 $\mu$Jy) in the 3.6, 4.5, 5.8, and 8.0~\um\ bands, respectively. The IRAC zeropoints are accurate at 
the 5\% level \citep{reach05} and the positions are accurate to 0\farcs3.

Since the X-ray survey covered a slightly larger area than the IR survey, we trimmed the X-ray survey by requiring that all the X-ray sources, including their error radii, were completely within the IR survey region. This resulted in a final X-ray catalog of 3086 sources.

\subsection{Mid-Infrared Identification of X-ray Sources}

Matching one catalog with $\sim$ 3100 sources, many of which have significant positional uncertainties, to a second
catalog with $\sim$ 270,000 sources presents some challenges.  If our matching criteria are too strict, then we 
have many false negatives, while if they are too liberal, we have many false positives.  

To best address this matching challenge we use the Bayesian source identification method used by \citet{brand06} for identifying the optical  counterparts to the XBo\"otes sources. The method optimizes the parameters of the matching criterion 
and supplies probabilities for both identifying the X-ray source with each nearby IRAC source and for 
the X-ray source having no identification in the IRAC catalog.  As a check on the Bayesian identification, 
we also used a simpler proximity match for the subset of 1658 X-ray sources with positional uncertainties $\leq$2\farcs0. 
This simple proximity match serves as a check for anomolies in the
results from the Bayesian method.  The results for both techniques and all four IRAC wavebands are listed 
in Table 1.   The fraction of identified sources depends
on the IR band because the sensitivity of the observations is greatest at 3.6 and 4.5 \um\ but
significantly worse at 5.8 and 8.0 \um. Figure~1 shows a comparison between the Bayesian and proximity match techniques based on the $[3.6]-[8.0]$ and $[5.8] - [8.0]$ colors and it is clear that the Bayesian matches are not introducing new populations, but the Bayesian technique is significantly increasing the number of matched sources in the bands where an increase is most useful. Since no problems were identified with the Bayesian method, we adopt
the Bayesian results for the remainder of our discussion. 

In terms of numbers of X-ray sources identified in the IR, we focus on the 3.6\um\ identifications as this is the most sensitive IRAC band.  
In the Bayesian method, each possible identification has a probability of being the correct 
identification. Figure~2 summarizes the distribution of the match probabilities for the
most likely 3.6 \um\ counterpart for each X-ray source.  Most sources (79\%) are matched 
at a confidence level above 95\%; 85\% are matched at a confidence level above 80\%.
In general, the sources with low match probabilities are relatively faint X-ray sources that are observed far off axis and have faint IRAC counterparts.

There are 349 objects (11\%) for which the highest likelihood is that
the X-ray source has no counterpart in the IRAC catalogs. We extracted
3.6 \um\ images of the regions near each of these objects for inspection.  In
most cases (244 objects) there is an IRAC source on or near the X-ray
position that is fainter than the $5\sigma$ limit of the
catalogs used for the matching procedure\footnote{Note that the 5$\sigma$ limit for the 6\arcs\ diameter aperture used for the IRAC catalog is a factor of 2 higher than the 5$\sigma$ limit for a 3\arcs\ aperture which more nearly corresponds to the {\it detection} limit of the survey \citep{eisenhardt04}. We use the 6\arcs\ apertures here to obtain more reliable colors.}.  For another 27 sources the X-ray source is in a region with either multiple, blended sources or very
close to a bright star.  Only
79 sources, 2.5\% of the total sample, show no obvious IRAC source
near the position of the X-ray source.  
Since the X-ray catalogs are
expected to have a roughly 1\% false detection rate, this means that
only $\sim 1.5\%$ of the true XBo\"otes sources have no counterpart
in the IRAC Shallow Survey.  Of these 79 sources, 70 have probable 
optical counterparts. However \citet{brand06} estimated
from Monte Carlo simulations that roughly 50\% of sources genuinely
lacking an optical counterpart would have a spurious counterpart due to the very
high optical source density.

IRAC is considerably less sensitive at 5.8 and 8.0~\um\ than at 3.6 and
4.5 \um, which could lead to biases in the IRAC color-color
distributions we will discuss shortly. We consider such biases by exploring how $[3.6]-[4.5]$ colors depend on the presence of 
5.8 and 8.0 \um\ detections.  Figure~3 shows
histograms of the $[3.6]-[4.5]$ color for (1) all matched objects with $>5\sigma$ detections in the blue channels (3.6 and 4.5\um) (2) $>5\sigma$ detections in all four channels, and (3) lacking $>5\sigma$ detections in the two
red channels (5.8 \& 8.0\um).  The median $[3.6]-[4.5]$ color of the
distributions are mutually consistent for all three cases, with median $[3.6]-[4.5]=0.49$, $0.56$ and $0.44$ for cases
(1), (2) and (3) respectively.  Objects with detections in all four
bands are marginally redder than the other cases, but the differences
are small enough to rule out a significant population of X-ray sources
with very blue $[3.6]-[4.5]$ colors which might be systematically missed at 5.8 and 8.0~\um.

\section{Infrared Properties of X-ray detected AGN}

Figure~4 shows the $[3.6]-[4.5]$ versus $[5.8]-[8.0]$ color-color
distribution of the X-ray sources.   The largest grouping of points running towards red $[3.6]-
[4.5]$ colors is the population of AGN identified by \citet{eisenhardt04} 
and also discussed in \citet{stern05}.  Superposed on Fig.~4 is the color selection criterion developed by \citet{stern05}
to identify AGN based on optical spectroscopy of 681 $R_{Vega}<$21.5~mag
AGN found in the survey region by AGES.  In the \citet{stern05}
spectroscopic sample, 77\% of the objects meeting the selection
criterion were broad line ({\it type I}) AGN, 6\% were narrow line AGN
({\it type II}) and 17\% were galaxies with no obvious signs of AGN activity
in their spectra.  These objects may also be AGN with optical
emission lines that are overwhelmed by the light of the
host galaxy \citep[e.g.][]{moran02} or absorbed by dust and gas \citep[e.g.][]{barger01}.
As shown in Fig.~4, of the 1325 X-ray sources with detections in all four IRAC bands, 65\% meet the \citet{stern05}
selection criterion.  

The problem is categorizing the remaining sources.  In Fig.~4 we have
divided the  $[3.6]-[4.5]$ versus $[5.8]-[8.0]$ color space into five
regions, where region A is the \citet{stern05} AGN selection region where power law emission of AGN from the near to the mid-IR dominates the flux of the host galaxy.
For each region we have also calculated the average X-ray hardness ratios
$HR=(H-S)/(H+S)$ of the sources, where $H$ is the number of hard
(2.0-7.0~keV) counts and S is the number of soft (0.5-2.0~keV)
counts.  In region A the median (average) source has $S= 6.00\pm2.45$ ($11.13\pm2.92$) and $H= 3.00\pm1.73$ ($4.95\pm1.95$) for a hardness ratio of $HR=-0.40\pm0.39$ ($-0.32\pm0.43$) (Figure~5). This value is typical of {\it type I} AGN \citep[e.g.][]{akylas04} implying the X-ray objects in this the region are dominated by {\it type I} sources as also evidenced by the optical spectroscopy \footnote{Typically {\it type I} AGN which have broad optical Hydrogen emission lines have low X-ray absorption leading to a softer X-ray hardness ratio, and {\it type II} AGN which have narrow optical Hydrogen emission lines have greater X-ray absorption, in particular of soft X-ray photons, which leads to a harder X-ray hardness ratio. These ratios are somewhat modified by redshift where the harder sources seem softer as the higher energy X-ray photons are redshifted to lower energies.} . A significant number of {\it Spitzer} selected sources also have well determined IRAC colors placing them in this AGN wedge but are not detected by XBo\"otes and are optically fainter than the AGES spectroscopic limit. Such sources are likely a mixture of fainter {\it type I} AGN and obscured, {\it type II}, AGN. Additionally there are X-ray sources which were detected off axis which are difficult to match to faint IRAC sources leading to the possibility that faint, obscured AGN which are more likely to be {\it type II} at all wavelengths \citep{ueda03,hao05,brown06,gorjian04} may be missed, but this is not the case. For Region A, if we take the subsample of sources with $>$2\arcs\  error radii which comprise 30\% of the sources in Region A, the hardness ratio changes only a small amount from -0.4 to -0.33. So a large number of faint IRAC {\it type II} sources are not being missed because they were preferentially selected against in off axis matching.

Region B, with 3\% of the sources, has a  median (average) hardness ratio of $HR=-0.24\pm0.45$ ($-0.25\pm0.47$) which is harder than that of region A. Figure 5 illustrates that the HR distribution extends towards harder values than for region A. {\it Type II} AGN have harder ratios than {\it type I} AGN since in obscured sources soft X-rays tend to be absorbed while hard X-rays are able to escape. So a mixture with relatively more {\it type II} AGN in region B than A would explain the difference in the shapes of the $HR$ distributions. In terms of the IR colors \citet{stern05} placed the blue limit on region A in $[5.8]-[8.0]$ color to avoid the inclusion of  $z > 1$ normal and starburst galaxies, but low luminosity AGN will have the same colors as normal galaxies \citep[e.g.][]{gorjian04}. So, in addition to a larger percentage of {\it type II}s,  region B is likely a combination of low IR luminosity AGN  at $z > 1$, as well the extension of AGN from region A.

Region C has a median (average) hardness ratio of $HR=-0.33\pm0.46$ ($-0.22\pm0.48$) containing 21\% of the sources.
Most of these sources are relatively blue in $[3.6]-[4.5]$ either because they are lower luminosity AGN with significant contamination of the mid-IR 
fluxes by their host galaxies or because of strong emission 
lines in the 3.6 \um\ band with Paschen $\alpha$ at $z \sim 0.8$ being the 
most important numerically, but also with contributions from H$\alpha$ at $z\sim 4.3$ \citep[e.g.][]{richards06}. The shape and value of the $HR$ distribution is similar to region A indicating a similar mix of {\it type I} and {\it II} AGN.

Region D contains objects with red $[5.8]-[8.0]$ colors indicative of low redshift
PAH emission associated with star formation and blue  $[3.6]-[4.5]$ colors indicative of the Rayleigh-Jeans fall-off of low-redshift stellar emission.  
This region contains 6\% of the sources and has a median (average) hardness
ratio of $HR=-0.20\pm0.47$ ($-0.06\pm0.48$). The difficulty lies in distinguishing emission
due to star formation from that due to obscured AGN, which can
be difficult even with spectroscopy.  The XBo\"otes survey is
deep enough to detect starbursts at modest redshifts ($z \simeq 0.04$
to $0.12$) so it is not unreasonable to expect a population of starbursts in the survey.\footnote{ In the models of \citet{leitherer95}, a region with a star formation rate of 1 $M_\sun$ yr$^{-1}$,  
a Salpeter initial mass function with a slope of 2.35, an upper mass cutoff of 100 $M_\sun$, 
and solar metallicity, produces 2.5\xx10$^4$ O stars over a period of 10$^7$ yr. 
Since the 2-10 keV X-ray luminosity from each O star is estimated to be between 2 and 
20\xx10$^{34}$ erg s$^{-1}$ \citep{helfand01}, a starburst forming $100 M_\odot$~year$^{-1}$
stars has a hard X-ray luminosity of $\sim$5-50\xx10$^{40}$ erg s$^{-1}$. That is detectable
to $z \simeq 0.04$ to $0.12$ given the hard X-ray sensitivity of the XBo\"otes survey  
(1.5\xx10$^{-14}$ erg cm$^{-2}$s$^{-1}$ \citet{kenter05}).} Starbursts, like {\it type II} AGN, tend to have hard X-ray spectra \citep{ptak97} . Since the median hardness ratio in region D is harder than that of the median source in region A, we expect that region D contains a greater mix of {\it type II} AGN and starbursts, but with a significant fraction of {\it type I} AGNs as evidenced by the presence of sources with soft X-ray hardness ratios. These are presumably lower luminosity {\it type I} AGN with weak nuclear IR emission that is not dominating the mid-IR colors and pushing them towards region A.

Finally, region E, the clump containing 4\% of the sources near
Vega colors of zero, is dominated by X-ray emitting Galactic stars and $z\sim0$ galaxies. Visual inspection of the NDWFS optical data shows that $\sim 20\%$ 
are large, extended, early type galaxies whose X-ray emission is likely from hot X-ray emitting gas rather than an AGN.  These sources
have softer spectra as compared to all the other regions with a median (average) $HR=-0.81\pm0.51$ ($-0.57\pm0.57$).

\subsection{A Smaller but Deeper Look: The LALA Survey}

A further check of these results comes from the single $17' \times 17'$ 
{\it Chandra}/ACIS field obtained by the Large Area Lyman Alpha Survey \citep[LALA,][]{rhoads00} 
with an  exposure time of 172~ksec \citep{wang04}.  These data reach a hard X-ray (2--10~keV) detection 
limit of $10^{-15}$~erg~cm$^{-2}$~s$^{-1}$, far deeper than the XBo\"otes survey limit of 1.5\xx$10^{-14}$~erg~cm$^{-2}$~s$^{-1}$.  We used a simple proximity match 
with a 2\farcs0 matching radius for the 168 cataloged sources since the median X-ray positional uncertainty
is only $0\farcs6$.  Of the 168 sources in the catalog, 130 (77\%) have
($5\sigma$) $3.6\mu$m counterparts within a $2\farcs0$ radius.  If we allow for visual matches to $3.6\mu$m sources fainter than
the $5\sigma$ catalog limit, then we find 145 (86\%) counterparts. This is a similar statistic to that  found by \citet{barmby06} from deep IRAC and {\it Chandra} imaging of a  $\sim20' \times 20'$ region of the Extended Groth Strip which found IRAC counterparts for 91\% of {\it Chandra} sources.  The high rate of recovery of the 172 ksec X-ray sources in the 90 sec IRAC Shallow Survey shows that the $\sim$1\% of the 5ksec sources that lack IRAC counterparts are not due simply to the fact that the IR survey is too shallow, but that the missing sources have some unusual X-Ray/IR characteristics.

In the LALA region, the XBo\"otes catalog has only 22 sources, of which 21 (95\%) have
3.6 \um\ counterparts.  The XBo\"otes sources in the LALA region are typical
of the field as a whole, with 64\% (9 of 14) of the 4-band detected sources satisfying the \citet{stern05} 
AGN selection criterion (Region A).  

Of the 35 sources in the deeper LALA image that are 
detected in all four channels, 17 (49\%) satisfy the \citet{stern05} AGN selection 
criterion and lie in region A. The majority of the rest, 13 (37\%), lie in region C.

With such small numbers it is difficult to make a comparison between the deep and shallow XBo\"otes source fractions, but the greater percentage of sources in region C  in the deeper survey may indicate a trend towards a larger fraction of obscured  AGN at fainter X-ray fluxes.

\section{Summary and Conclusions}

In this paper we examine the mid-infrared properties of 3086 X-ray
sources in the {\it Chandra} XBo\"otes Survey \citep{kenter05} that were detected by the
IRAC Shallow Survey \citep{eisenhardt04} -- the largest comparison of X-ray and mid-
infrared sources yet undertaken.  Despite an integration time of only
90~sec, the IRAC Shallow Survey detects 85\% of the X-ray sources, with
another 13\% being detectable at lower confidence levels than the $5
\sigma$ detection limit of the primary Shallow Survey catalogs.  Only 2.5\% of the X-ray sources, up to 40\% of which may represent false-positives in the X-ray catalogs, lack a counterpart.   Even in the small
area but deeper LALA X-ray survey, based on an X-ray exposure
time of 172~ksec rather than 5~ksec, most of the X-ray sources are
easily detected (77\% are in the $5\sigma$ catalogs and 86\% are
detected at deeper thresholds).

The mid-infrared colors of the X-ray sources show five relatively
distinct classes.  By far the largest class of sources (65\%) satisfy the simple color-selection criteria
developed by \citet{stern05} to select AGN in the mid-infrared (Region A).
Most of the remaining sources lie in an extension of this region where redshifted emission lines and/or host galaxy contributions to the SED provide for bluer $[3.6]-[4.5]$ colors (Regions B and C).  Sources with red $[5.8]-[8.0]$ colors and blue $[3.6]-[4.5]$ colors are likely dominated by obscured AGN, lower luminosity unobscured AGN, and starburst galaxies (Region D). Finally, small fractions of the sources are
clustered near mid-IR colors of zero magnitude (Region E).  These sources are a
combination of X-ray emitting stars and  low redshift galaxies
whose X-ray flux comes from X-ray emitting hot gas.

This segregation in color space makes the mid-IR a very efficient wavelength range to do large area surveys for luminous IR AGN.  In comparison to the XBo\"otes survey which detected $\sim$3000 sources in 630 ks, the IRAC Shallow Survey detected $\sim$2000 AGN in 216 ks ($\sim$2000 is the {\it total} \citet{stern05} number of IRAC objects in the wedge), a factor of 2 higher in detection efficiency. The X-rays though can help provide a completeness factor for lower luminosity AGN which lie outside the wedge. \citet{stern05} deduced a lower limit surface density of 250 AGN per deg$^2$ based on the number of sources in region A from the entire Shallow Survey data (corrected by 9\% to 275 per deg$^2$ based on spectroscopy of sources outside the wedge). The X-rays allow us to increase this lower limit by adding the AGN from regions B and C, which contain an additional 25\% of four band detected sources, raising the lower limit to 350 AGN per deg$^2$. 

Another approach is to directly combine the two techniques. With the exception for the stars in region E and the possible contamination of luminous starbursts in region D, 90\% of the X-ray sources with 4 channel IRAC detections are AGN accounting for $\sim2700$ sources from the full X-ray catalog. Of these X-ray sources 864 had 4 channel IRAC detections placing them in region A. Of the IRAC sources there are $\sim2000$ sources in region A. Combining all the sources detected in the IR in region A with all the X-ray sources which are AGN  (making sure not to double count the 864 IR detected X-ray sources in region A), gives a total of $\sim4000$ AGN for a surface density of 460 AGN per deg$^2$. Thus, using either approach can help place better limits on the contribution of AGN to the CIB.

%We analyze the mid-IR photometry from the IRAC Shallow Survey of 3086 X-ray detected AGN from the {\it Chandra} XBo\"otes survey. This comparison is the largest in number and area yet undertaken. We find that in relatively short exposure times, 90 sec for {\it Spitzer} and 5 ks for {\it Chandra}, that 85\% of the AGN detected in the X-ray have mid-IR counterparts down to a confidence level of 80\%. Of the remaining 15\%, 13\% were below the 5$\sigma$ detection threshhold and 2\% did not have a counterpart. Based on the 1\% spurious rate cutoff in the X-rays, out of the 2\% with no detections, 1\% likely fell below the 3.6~\um\ sensitivity.

%The mid-IR colors of the majority of AGN (67\%) have a distinct grouping in color-color space which match those empirically defined from optical spectroscopy.  A small number of objects near zero-zero colors (5\%) are likely foreground stars or nearby E-S0 type galaxies whose X-ray flux comes from X-ray binaries or hot gas.

%The good match between the surveys shows that both the infrared and X-ray detected objects come from the same population and a combined approach of X-ray and mid-IR photometry is a powerful tool in surveying for AGN and then determining the contribution of accretion processes to the CIB.

\acknowledgments
This work is based on data from two facilities: the {\it Spitzer Space Telescope}, which is operated by the Jet Propulsion Laboratory, California Institute of Technology under contract with NASA and the {\it Chandra X-ray Observatory}, which is operated by the Harvard Smithsonian Astrophysical Observatory. Support for this work was provided by NASA. The NOAO is operated by the Association of Universities for Research in Astronomy (AURA) Inc. under cooperative agreement with the National Science Foundation. This research has made use of the NASA/IPAC Extragalactic Database (NED) which is operated by the Jet Propulsion Laboratory, California Institute of Technology, under contract with the National Aeronautics and Space Administration.

%\clearpage
%\bibliographystyle{apj}
%\bibliography{apj-jour,irac-cxo.bib}
%\bibliography{a/irac-cxo.bib}

\begin{thebibliography}{30}
\expandafter\ifx\csname natexlab\endcsname\relax\def\natexlab#1{#1}\fi

\bibitem[{{Akylas} {et~al.}(2004){Akylas}, {Georgakakis}, \&
  {Georgantopoulos}}]{akylas04}
{Akylas}, A., {Georgakakis}, A., \& {Georgantopoulos}, I. 2004, \mnras, 353,
  1015

\bibitem[{{Alonso-Herrero} {et~al.}(2004){Alonso-Herrero},
  {P{\'e}rez-Gonz{\'a}lez}, {Rigby}, {Rieke}, {Le Floc'h}, {Barmby}, {Page},
  {Papovich}, {Dole}, {Egami}, {Huang}, {Rigopoulou},
  {Crist{\'o}bal-Hornillos}, {Eliche-Moral}, {Balcells}, {Prieto}, {Erwin},
  {Engelbracht}, {Gordon}, {Werner}, {Willner}, {Fazio}, {Frayer}, {Hines},
  {Kelly}, {Latter}, {Misselt}, {Miyazaki}, {Morrison}, {Rieke}, \&
  {Wilson}}]{alonso04}
{Alonso-Herrero}, A., {P{\'e}rez-Gonz{\'a}lez}, P.~G., {Rigby}, J., {Rieke},
  G.~H., {Le Floc'h}, E., {Barmby}, P., {Page}, M.~J., {Papovich}, C., {Dole},
  H., {Egami}, E., {Huang}, J.-S., {Rigopoulou}, D., {Crist{\'o}bal-Hornillos},
  D., {Eliche-Moral}, C., {Balcells}, M., {Prieto}, M., {Erwin}, P.,
  {Engelbracht}, C.~W., {Gordon}, K.~D., {Werner}, M., {Willner}, S.~P.,
  {Fazio}, G.~G., {Frayer}, D., {Hines}, D., {Kelly}, D., {Latter}, W.,
  {Misselt}, K., {Miyazaki}, S., {Morrison}, J., {Rieke}, M.~J., \& {Wilson},
  G. 2004, \apjs, 154, 155

\bibitem[{{Barger} {et~al.}(2001){Barger}, {Cowie}, {Mushotzky}, \&
  {Richards}}]{barger01}
{Barger}, A.~J., {Cowie}, L.~L., {Mushotzky}, R.~F., \& {Richards}, E.~A. 2001,
  \aj, 121, 662

\bibitem[{{Barmby et~al.}(2006)}]{barmby06}
{Barmby et~al.} 2006, \apj, 642, 126

\bibitem[{{Bertin} \& {Arnouts}(1996)}]{bertin96}
{Bertin}, E. \& {Arnouts}, S. 1996, \aaps, 117, 393

\bibitem[{{Brand} {et~al.}(2006){Brand}, {Eisenhardt}, {Gorjian}, {Kochanek},
  {Caldwell}, {Eisenstein}, {Brodwin}, {Brown}, {Cool}, {Dey}, {Green},
  {Jannuzi}, {Murray}, {Pahre}, \& {Willner}}]{brand06}
{Brand}, K., {Eisenhardt}, P., {Gorjian}, V., {Kochanek}, C.~S., {Caldwell},
  N., {Eisenstein}, D., {Brodwin}, M., {Brown}, M.~I., {Cool}, R., {Dey}, A.,
  {Green}, P., {Jannuzi}, B.~T., {Murray}, S.~S., {Pahre}, M.~A., \& {Willner},
  S.~P. 2006, \apj, 641, 140

\bibitem[{{Brown} {et~al.}(2006){Brown}, {Brand}, {Dey}, {Jannuzi}, {Cool}, {Le
  Floc'h}, {Kochanek}, {Armus}, {Bian}, {Higdon}, {Higdon}, {Papovich},
  {Rieke}, {Rieke}, {Smith}, {Soifer}, \& {Weedman}}]{brown06}
{Brown}, M.~J.~I., {Brand}, K., {Dey}, A., {Jannuzi}, B.~T., {Cool}, R., {Le
  Floc'h}, E., {Kochanek}, C.~S., {Armus}, L., {Bian}, C., {Higdon}, J.,
  {Higdon}, S., {Papovich}, C., {Rieke}, G., {Rieke}, M., {Smith}, J.~D.,
  {Soifer}, B.~T., \& {Weedman}, D. 2006, \apj, 638, 88

\bibitem[{{Devriendt} {et~al.}(1999){Devriendt}, {Guiderdoni}, \&
  {Sadat}}]{devriendt99}
{Devriendt}, J.~E.~G., {Guiderdoni}, B., \& {Sadat}, R. 1999, \aap, 350, 381

\bibitem[{{Eisenhardt} {et~al.}(2004){Eisenhardt}, {Stern}, {Brodwin}, {Fazio},
  {Rieke}, {Rieke}, {Werner}, {Wright}, {Allen}, {Arendt}, {Ashby}, {Barmby},
  {Forrest}, {Hora}, {Huang}, {Huchra}, {Pahre}, {Pipher}, {Reach}, {Smith},
  {Stauffer}, {Wang}, {Willner}, {Brown}, {Dey}, {Jannuzi}, \&
  {Tiede}}]{eisenhardt04}
{Eisenhardt}, P.~R., {Stern}, D., {Brodwin}, M., {Fazio}, G.~G., {Rieke},
  G.~H., {Rieke}, M.~J., {Werner}, M.~W., {Wright}, E.~L., {Allen}, L.~E.,
  {Arendt}, R.~G., {Ashby}, M.~L.~N., {Barmby}, P., {Forrest}, W.~J., {Hora},
  J.~L., {Huang}, J.-S., {Huchra}, J., {Pahre}, M.~A., {Pipher}, J.~L.,
  {Reach}, W.~T., {Smith}, H.~A., {Stauffer}, J.~R., {Wang}, Z., {Willner},
  S.~P., {Brown}, M.~J.~I., {Dey}, A., {Jannuzi}, B.~T., \& {Tiede}, G.~P.
  2004, \apjs, 154, 48

\bibitem[{{Fadda} {et~al.}(2002){Fadda}, {Flores}, {Hasinger}, {Franceschini},
  {Altieri}, {Cesarsky}, {Elbaz}, \& {Ferrando}}]{fadda02}
{Fadda}, D., {Flores}, H., {Hasinger}, G., {Franceschini}, A., {Altieri}, B.,
  {Cesarsky}, C.~J., {Elbaz}, D., \& {Ferrando}, P. 2002, \aap, 383, 838

\bibitem[{{Franceschini} {et~al.}(2005){Franceschini}, {Manners}, {Polletta},
  {Lonsdale}, {Gonzalez-Solares}, {Surace}, {Shupe}, {Fang}, {Xu}, {Farrah},
  {Berta}, {Rodighiero}, {Perez-Fournon}, {Hatziminaoglou}, {Smith}, {Siana},
  {Rowan-Robinson}, {Nandra}, {Babbedge}, {Vaccari}, {Oliver}, {Wilkes},
  {Owen}, {Padgett}, {Frayer}, {Jarrett}, {Masci}, {Stacey}, {Almaini},
  {McMahon}, {Johnson}, {Lawrence}, \& {Willott}}]{franceschini05}
{Franceschini}, A., {Manners}, J., {Polletta}, M.~d.~C., {Lonsdale}, C.,
  {Gonzalez-Solares}, E., {Surace}, J., {Shupe}, D., {Fang}, F., {Xu}, C.~K.,
  {Farrah}, D., {Berta}, S., {Rodighiero}, G., {Perez-Fournon}, I.,
  {Hatziminaoglou}, E., {Smith}, H.~E., {Siana}, B., {Rowan-Robinson}, M.,
  {Nandra}, K., {Babbedge}, T., {Vaccari}, M., {Oliver}, S., {Wilkes}, B.,
  {Owen}, F., {Padgett}, D., {Frayer}, D., {Jarrett}, T., {Masci}, F.,
  {Stacey}, G., {Almaini}, O., {McMahon}, R., {Johnson}, O., {Lawrence}, A., \&
  {Willott}, C. 2005, \aj, 129, 2074

\bibitem[{{Gorjian} {et~al.}(2004){Gorjian}, {Werner}, {Jarrett}, {Cole}, \&
  {Ressler}}]{gorjian04}
{Gorjian}, V., {Werner}, M.~W., {Jarrett}, T.~H., {Cole}, D.~M., \& {Ressler},
  M.~E. 2004, \apj, 605, 156

\bibitem[{{Hao} {et~al.}(2005){Hao}, {Strauss}, {Fan}, {Tremonti}, {Schlegel},
  {Heckman}, {Kauffmann}, {Blanton}, {Gunn}, {Hall}, {Ivezi{\'c}}, {Knapp},
  {Krolik}, {Lupton}, {Richards}, {Schneider}, {Strateva}, {Zakamska},
  {Brinkmann}, \& {Szokoly}}]{hao05}
{Hao}, L., {Strauss}, M.~A., {Fan}, X., {Tremonti}, C.~A., {Schlegel}, D.~J.,
  {Heckman}, T.~M., {Kauffmann}, G., {Blanton}, M.~R., {Gunn}, J.~E., {Hall},
  P.~B., {Ivezi{\'c}}, {\v Z}., {Knapp}, G.~R., {Krolik}, J.~H., {Lupton},
  R.~H., {Richards}, G.~T., {Schneider}, D.~P., {Strateva}, I.~V., {Zakamska},
  N.~L., {Brinkmann}, J., \& {Szokoly}, G.~P. 2005, \aj, 129, 1795

\bibitem[{{Hauser} \& {Dwek}(2001)}]{hauser01}
{Hauser}, M.~G. \& {Dwek}, E. 2001, \araa, 39, 249

\bibitem[{{Helfand} \& {Moran}(2001)}]{helfand01}
{Helfand}, D.~J. \& {Moran}, E.~C. 2001, \apj, 554, 27

\bibitem[{{Hickox} {et~al.}(2007){Hickox}, {Jones}, {Forman}, {Murray}, {Brodwin}, {Brown}, {Eisenhardt}, 	{Stern}, {Kochanek}, {Eisenstein}, {Cool}, 	{Jannuzi}, {Dey}, {Brand}, {Gorjian}, \& {Caldwell}}]{hickox07} {Hickox}, R.~C. and {Jones}, C. and {Forman}, W.~R. and {Murray}, S.~S. and 	{Brodwin}, M. and {Brown}, M.~J.~I. and {Eisenhardt}, P.~R. and 	{Stern}, D. and {Kochanek}, C.~S. and {Eisenstein}, D. and {Cool}, R.~J. and {Jannuzi}, B.~T. and {Dey}, A. and {Brand}, K. and {Gorjian}, V. and {Caldwell}, N. 2007, \apj, 671, 1365


\bibitem[{{Jannuzi} \& {Dey}(1999)}]{jannuzi99}
{Jannuzi}, B.~T. \& {Dey}, A. 1999, in ASP Conf. Ser. 193: The Hy-Redshift
  Universe: Galaxy Formation and Evolution at High Redshift, 258--+

\bibitem[{{Kenter} {et~al.}(2005){Kenter}, {Murray}, {Forman}, {Jones},
  {Green}, {Kochanek}, {Vikhlinin}, {Fabricant}, {Fazio}, {Brand}, {Brown},
  {Dey}, {Jannuzi}, {Najita}, {McNamara}, {Shields}, \& {Rieke}}]{kenter05}
{Kenter}, A., {Murray}, S.~S., {Forman}, W.~R., {Jones}, C., {Green}, P.,
  {Kochanek}, C.~S., {Vikhlinin}, A., {Fabricant}, D., {Fazio}, G., {Brand},
  K., {Brown}, M.~J.~I., {Dey}, A., {Jannuzi}, B.~T., {Najita}, J., {McNamara},
  B., {Shields}, J., \& {Rieke}, M. 2005, \apjs, 161, 9

\bibitem[{{Lacy} {et~al.}(2004){Lacy}, {Storrie-Lombardi}, {Sajina},
  {Appleton}, {Armus}, {Chapman}, {Choi}, {Fadda}, {Fang}, {Frayer},
  {Heinrichsen}, {Helou}, {Im}, {Marleau}, {Masci}, {Shupe}, {Soifer},
  {Surace}, {Teplitz}, {Wilson}, \& {Yan}}]{lacy04}
{Lacy}, M., {Storrie-Lombardi}, L.~J., {Sajina}, A., {Appleton}, P.~N.,
  {Armus}, L., {Chapman}, S.~C., {Choi}, P.~I., {Fadda}, D., {Fang}, F.,
  {Frayer}, D.~T., {Heinrichsen}, I., {Helou}, G., {Im}, M., {Marleau}, F.~R.,
  {Masci}, F., {Shupe}, D.~L., {Soifer}, B.~T., {Surace}, J., {Teplitz}, H.~I.,
  {Wilson}, G., \& {Yan}, L. 2004, \apjs, 154, 166

\bibitem[{{Leitherer} {et~al.}(1995){Leitherer}, {Robert}, \&
  {Heckman}}]{leitherer95}
{Leitherer}, C., {Robert}, C., \& {Heckman}, T.~M. 1995, \apjs, 99, 173

\bibitem[{{Moran} {et~al.}(2002){Moran}, {Filippenko}, \& {Chornock}}]{moran02}
{Moran}, E.~C., {Filippenko}, A.~V., \& {Chornock}, R. 2002, \apjl, 579, L71

\bibitem[{{Murray} {et~al.}(2005){Murray}, {Forman}, {Jones}, {Kenter},
  {Green}, {Fabricant}, {Fazio}, {Jannuzi}, {Dey}, {Najita}, {Brown}, {Brand},
  {Shields}, {McNamara}, {Rieke}, \& {Kochanek}}]{murray04}
{Murray}, S.~S., {Forman}, W.~R., {Jones}, C.~F., {Kenter}, A., {Green}, P.~J.,
  {Fabricant}, D.~G., {Fazio}, G.~G., {Jannuzi}, B.~T., {Dey}, A.~T., {Najita},
  J.~R., {Brown}, M.~J., {Brand}, K.~J., {Shields}, J.~C., {McNamara}, B.,
  {Rieke}, M.~J., \& {Kochanek}, C.~S. 2005, astro--ph/0504084

\bibitem[{{Ptak} {et~al.}(1997){Ptak}, {Serlemitsos}, {Yaqoob}, {Mushotzky}, \&
  {Tsuru}}]{ptak97}
{Ptak}, A., {Serlemitsos}, P., {Yaqoob}, T., {Mushotzky}, R., \& {Tsuru}, T.
  1997, \aj, 113, 1286

\bibitem[{{Reach} {et~al.}(2005){Reach}, {Megeath}, {Cohen}, {Hora}, {Carey},
  {Surace}, {Willner}, {Barmby}, {Wilson}, {Glaccum}, {Lowrance}, {Marengo}, \&
  {Fazio}}]{reach05}
{Reach}, W.~T., {Megeath}, S.~T., {Cohen}, M., {Hora}, J., {Carey}, S.,
  {Surace}, J., {Willner}, S.~P., {Barmby}, P., {Wilson}, G., {Glaccum}, W.,
  {Lowrance}, P., {Marengo}, M., \& {Fazio}, G.~G. 2005, \pasp, 117, 978

\bibitem[{{Rhoads} {et~al.}(2000){Rhoads}, {Malhotra}, {Dey}, {Stern},
  {Spinrad}, \& {Jannuzi}}]{rhoads00}
{Rhoads}, J.~E., {Malhotra}, S., {Dey}, A., {Stern}, D., {Spinrad}, H., \&
  {Jannuzi}, B.~T. 2000, \apjl, 545, L85

\bibitem[{{Richards et~al.}(2006)}]{richards06}
{Richards et~al.} 2006, \apjs, Submitted

\bibitem[{{Stern} {et~al.}(2005){Stern}, {Eisenhardt}, {Gorjian}, {Kochanek},
  {Caldwell}, {Eisenstein}, {Brodwin}, {Brown}, {Cool}, {Dey}, {Green},
  {Jannuzi}, {Murray}, {Pahre}, \& {Willner}}]{stern05}
{Stern}, D., {Eisenhardt}, P., {Gorjian}, V., {Kochanek}, C.~S., {Caldwell},
  N., {Eisenstein}, D., {Brodwin}, M., {Brown}, M.~J.~I., {Cool}, R., {Dey},
  A., {Green}, P., {Jannuzi}, B.~T., {Murray}, S.~S., {Pahre}, M.~A., \&
  {Willner}, S.~P. 2005, \apj, 631, 163

\bibitem[{{Ueda} {et~al.}(2003){Ueda}, {Akiyama}, {Ohta}, \& {Miyaji}}]{ueda03}
{Ueda}, Y., {Akiyama}, M., {Ohta}, K., \& {Miyaji}, T. 2003, \apj, 598, 886

\bibitem[{{Wang} {et~al.}(2004){Wang}, {Malhotra}, {Rhoads}, {Brown}, {Dey},
  {Heckman}, {Jannuzi}, {Norman}, {Tiede}, \& {Tozzi}}]{wang04}
{Wang}, J.~X., {Malhotra}, S., {Rhoads}, J.~E., {Brown}, M.~J.~I., {Dey}, A.,
  {Heckman}, T.~M., {Jannuzi}, B.~T., {Norman}, C.~A., {Tiede}, G.~P., \&
  {Tozzi}, P. 2004, \aj, 127, 213

\bibitem[{{Weisskopf} {et~al.}(1996){Weisskopf}, {O'dell}, \& {van
  Speybroeck}}]{weisskopf96}
{Weisskopf}, M.~C., {O'dell}, S.~L., \& {van Speybroeck}, L.~P. 1996, in Proc.
  SPIE Vol. 2805, p. 2-7, Multilayer and Grazing Incidence X-Ray/EUV Optics
  III, Richard B. Hoover; Arthur B. Walker; Eds., 2--7

\bibitem[{{Werner} {et~al.}(2004){Werner}, {Roellig}, {Low}, {Rieke}, {Rieke},
  {Hoffmann}, {Young}, {Houck}, {Brandl}, {Fazio}, {Hora}, {Gehrz}, {Helou},
  {Soifer}, {Stauffer}, {Keene}, {Eisenhardt}, {Gallagher}, {Gautier}, {Irace},
  {Lawrence}, {Simmons}, {Van Cleve}, {Jura}, {Wright}, \&
  {Cruikshank}}]{werner04}
{Werner}, M.~W., {Roellig}, T.~L., {Low}, F.~J., {Rieke}, G.~H., {Rieke}, M.,
  {Hoffmann}, W.~F., {Young}, E., {Houck}, J.~R., {Brandl}, B., {Fazio}, G.~G.,
  {Hora}, J.~L., {Gehrz}, R.~D., {Helou}, G., {Soifer}, B.~T., {Stauffer}, J.,
  {Keene}, J., {Eisenhardt}, P., {Gallagher}, D., {Gautier}, T.~N., {Irace},
  W., {Lawrence}, C.~R., {Simmons}, L., {Van Cleve}, J.~E., {Jura}, M.,
  {Wright}, E.~L., \& {Cruikshank}, D.~P. 2004, \apjs, 154, 1

\end{thebibliography}
%\end{thebibliography}

\clearpage

\begin{deluxetable}{lcccccccc}
%\rotate
\tablewidth{0pt}
%\tabletypesize{\scriptsize}
\setlength{\tabcolsep}{0.08in}
\tablecaption{Numbers (Percent) of XBo\"otes Sources Identified in the Mid-Infrared}

\tablehead{
\colhead{} &
\colhead{3.6 \um} &
%\colhead{3.6 \um} &
\colhead{4.5 \um} &
%\colhead{4.5 \um} &
\colhead{5.8 \um} &
%\colhead{5.8 \um} &
\colhead{8.0 \um} &
%\colhead{8.0 \um} \\

%\colhead{} &
%\colhead{$>$80\%$^1$} &
%\colhead{$\leq$ 2\arcs$^2$} &
%\colhead{$>$80\%} &
%\colhead{$\leq$ 2\arcs} &
%\colhead{$>$80\%} &
%\colhead{$\leq$ 2\arcs} &
%\colhead{$>$80\%} &
%\colhead{$\leq$ 2\arcs} 
}

\startdata

Bayesian matching$^1$  & 2609 (85\%) & 2422 (78\%) & 1346 (44\%) & 1487 (48\%) \\ 
Proximity matching$^2$  & 1477 (48\%) & 1400 (45\%)  & 881 (29\%) & 945 (31\%) \\

\enddata

\tablenotetext{1}{Matching is done using the Bayesian method with an 80\% threshhold in the Bayesian confidence level.}
\tablenotetext{2}{Matching is done by identifying a mid-IR source that is within 2\farcs0 of an X-ray source which has a positonal accuracy of $\leq$2\farcs0.}
\tablecomments{These are proximity match percentages based on the full 3086 X-ray sample. If one were only to compare the proximity matches with the 1658 sources that have positional errors $\leq$2\farcs0, then the statistics would be the following: 3.6\um=89\%, 4.5\um=84\%, 5.8\um=54\%, 8.0\um=57\%. }

\end{deluxetable}

\clearpage

\begin{deluxetable}{lrrr}
%\rotate
\tablewidth{0pt}
%\tabletypesize{\scriptsize}
\setlength{\tabcolsep}{0.08in}
\tablecaption{Summary of Statistics}

\tablehead{
\colhead{} &
\colhead{} &
\colhead{Percentage of all} &
\colhead {Percentage of sources} \\

\colhead{} &
\colhead{Number} &
\colhead{3086 sources} &
\colhead{detected in all}\\

\colhead{} &
\colhead{} &
\colhead{} &
\colhead{4 IRAC bands}

}

\startdata

%3.6\um\ sources & $\sim$270,000 \\ 

X-ray sources & 3086 & 100 & -\\ 

----- with 3.6\um\  counterparts$^1$ & 2609 & 85 & - \\ 

----- with 4 color mid-IR counterparts  & 1325 & 42 & 100\\ 
 
-----  Region A:  Within AGN wedge & 864 & 28  & 65\\

----- Region B: Blue $[5.8]-[8.0]$ colors  & 46 & 2 & 4\\

----- Region C: Red extension of region A & 283 & 9 & 21\\

----- Region D: Red $[5.8]-[8.0]$ color & 78  & 3 & 6\\

----- Region E: Near zero-zero color & 57  & 2 & 4\\

%Total number of 3.6\um\ sources & $\sim$270,000 \\ 

%Total number of 3.6\um\ sources & $\sim$270,000 \\ 

\enddata

\tablenotetext{1}{The counterparts are all sources with $>$80\% Bayesian probability of  a mid-IR source matching an X-ray source.}
\tablecomments{Regions A-E are the regions identified in Figure 4}

\end{deluxetable}

\clearpage

\begin{figure}
\epsscale{0.80}
\plotone{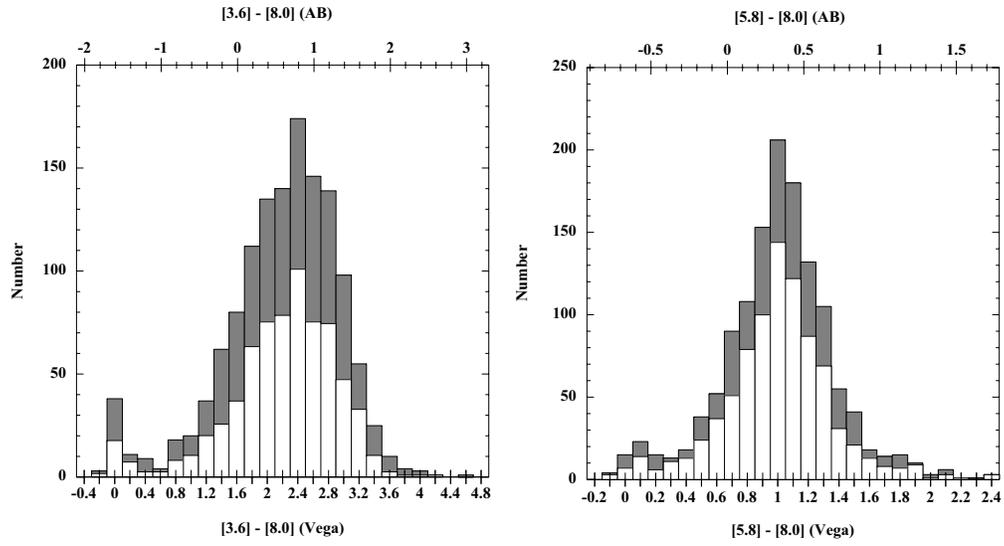}
\figcaption{A comparison of the $[3.6]-[8.0]$ (left) and the $[5.8]-[8.0] $ (right) colors of the sources identified with the proximity match technique (white) and the Bayesian technique (grey). Note that the Bayesian technique identifies significantly more sources without introducing new populations when comparing color baselines that are large ([3.6]-[8.0]) and small ([5.8]-[8.0]).  \label{bayes-vs-prox}}
\end{figure}

\clearpage

\begin{figure}
\epsscale{0.80}
\plotone{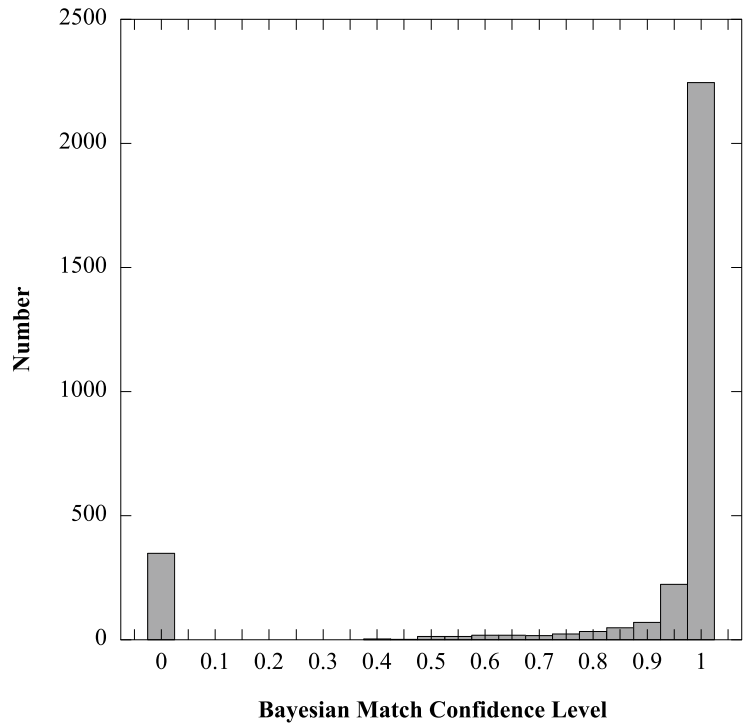}
\figcaption{The Bayesian match statistics results for 3.6\um\ identifications of XBo\"otes sources. 79\% of the X-ray objects have IR matches at the 95\% confidence level or greater. 85\% of the X-ray objects have matches at 80\% confidence level and above.
\label{stats_01}}
\end{figure}

\clearpage

\begin{figure}
%\epsscale{0.75}
\plotone{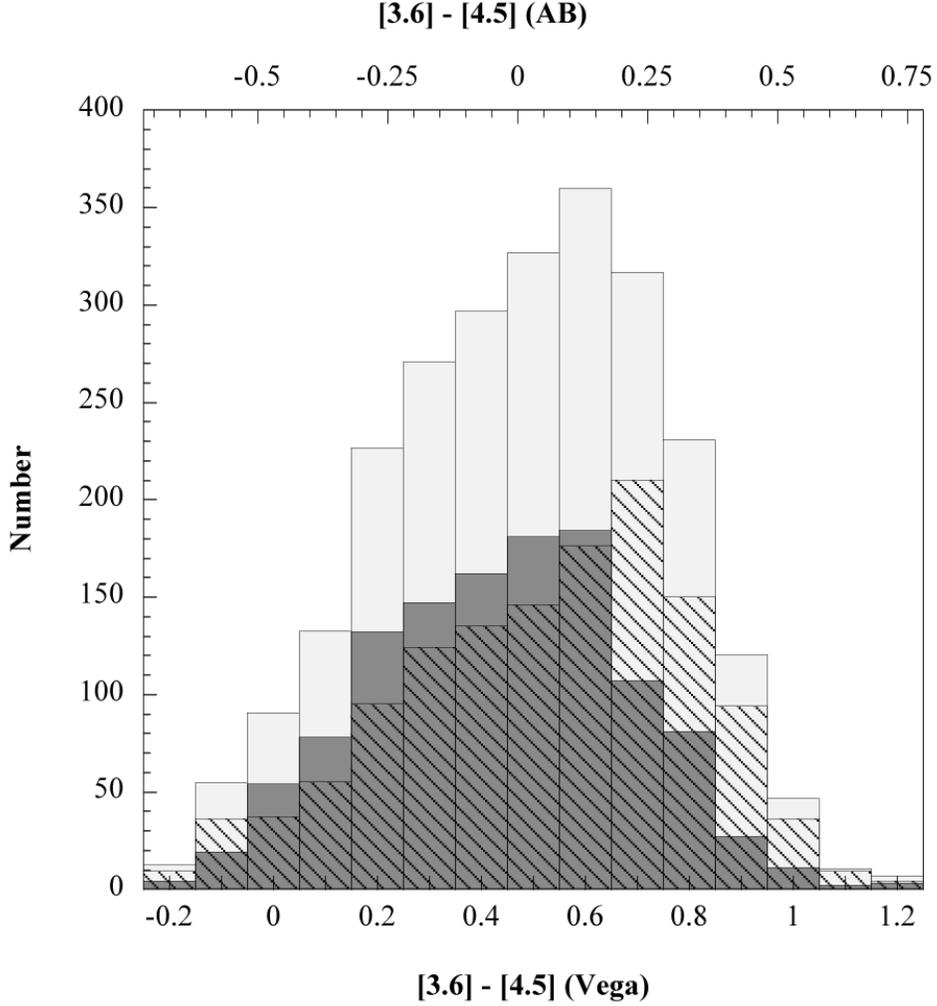}
\figcaption{Histogram of [3.6] - [4.5] colors for three sets of the X-ray detected sample. The light grey histogram is for all matched objects with $>5\sigma$ detections in the blue channels (3.6 and 4.5\um) (median 0.49). The hatched histogram plots the [3.6] - [4.5] colors for objects with $>$5$\sigma$ in all four channels (median 0.56). The dark grey histogram plots those objects lacking $>$5$\sigma$ detections in the two red channels (5.8 \& 8.0\um) (median 0.44). Note that although objects with $>$5$\sigma$ detections in all four channels are slightly redder in [3.6] - [4.5] color than those without 5.8 and 8.0~\um\ detections, the blue side of the distributions are similar in shape and numbers so that there is not a significant population of X-ray objects with very blue [3.6] - [4.5] colors that is being missed.
\label{mips_01} }
\end{figure}

\clearpage

\begin{figure}
\epsscale{0.80}
\plotone{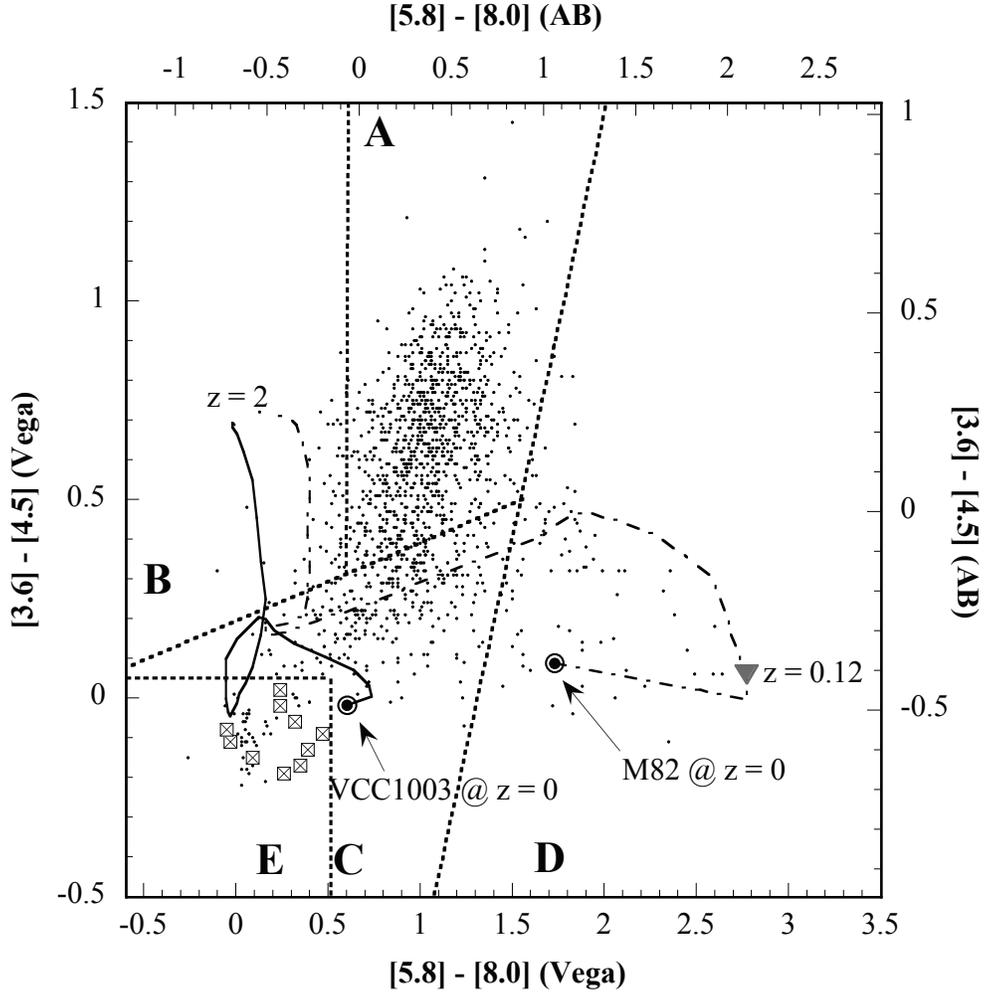}
\figcaption{Mid-infrared color-color diagram for all 1325 XBo\"otes detections with $>$5$\sigma$ detections in all four IRAC bands. Region A is  the empirically defined region of AGN based on optical spectroscopy from \citet{stern05}. Region B has less IR luminous AGN at $z > 1$ as shown by the color track from z = 0 to 2 of the S0/Sa galaxy VCC1003=NGC4429 based on a template from \citet{devriendt99}. This region also contains the extension of region A which was excluded in the optical due to the possibility of redshifted starburst galaxies as shown by the M82 color track which has a star formation rate $\sim4000$ times higher than VCC1003. Region C is the extension of region A representing less luminous AGN and sources at redshifts with prominent emission lines in the 3.6 \um\ bandpass. Region D likely is dominated by lower luminosity AGN and starbursts. Due to the limited sensitivity of the XBo\"otes survey, the highest redshift starburst galaxy will not likely be detected beyond a z = 0.12 noted by the triangle. Region E is a combination of X-ray emitting stars in our own galaxy and optically extended sources (noted as crossed boxes), presumably from nearby (non-AGN) E/S0 galaxies with bright X-ray emitting gas. \label{mips_02} }
\end{figure}

\clearpage

\begin{figure}
\epsscale{0.9}
\plotone{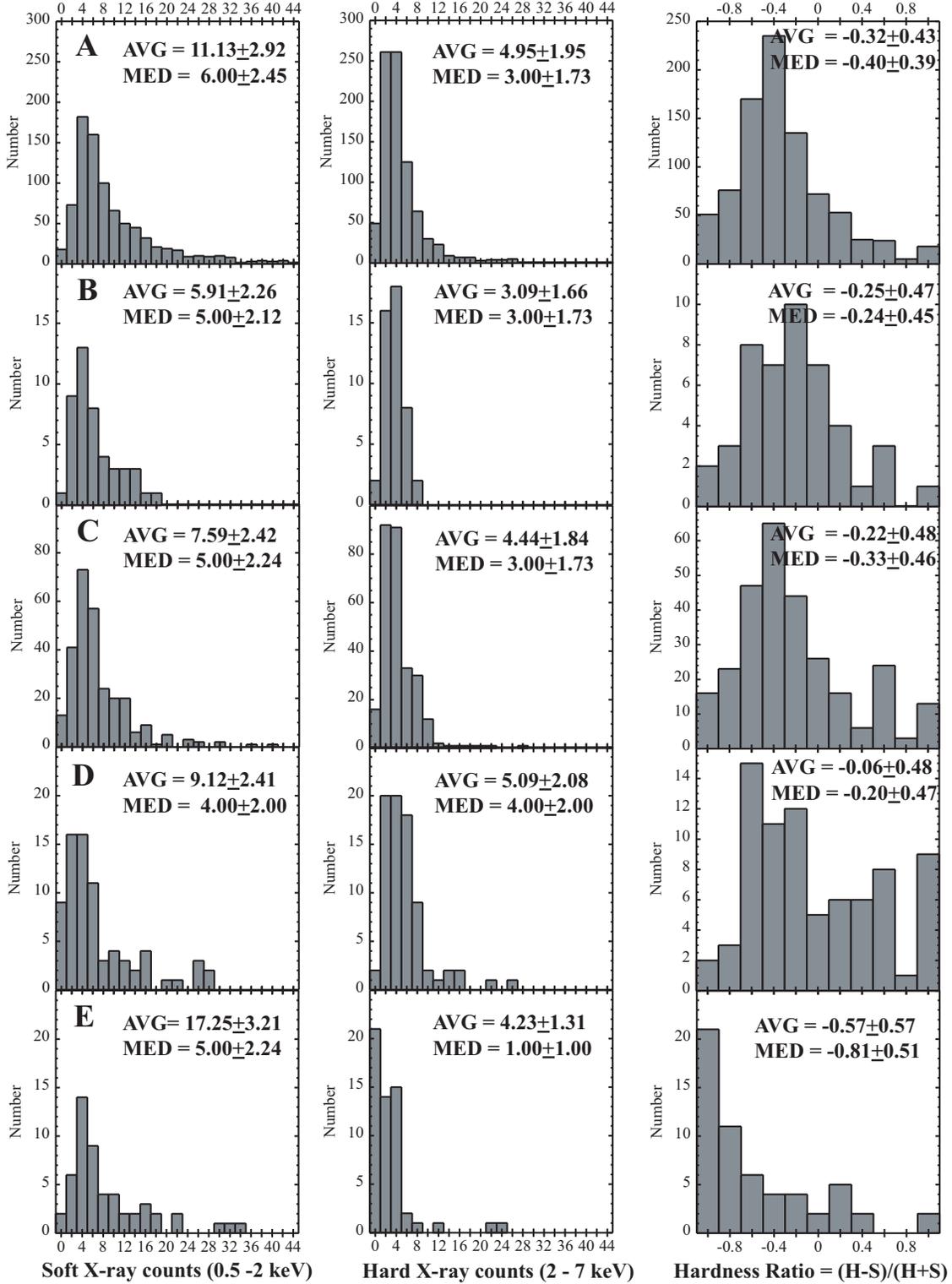}
\figcaption{Distributions of soft (0.5 - 2 keV) X-ray counts, hard (2 - 7 keV) X-ray counts, and Hardness Ratios = (H-S)/(H+S) for XBo\"otes sources in five regions of mid-IR color-color space. (See Figure 4).
\label{stats_02}}
\end{figure}

\end{document}